\documentclass[sigconf]{acmart}

\usepackage{amsmath,amsfonts,amsthm}
\usepackage{algorithmic}
\usepackage{algorithm}
\usepackage{graphicx}
\usepackage{textcomp}
\usepackage{xcolor}
\usepackage{booktabs}
\usepackage{listings}
\usepackage{tabularx}
\usepackage{multirow}
\usepackage{url}
\usepackage{hyperref}
\usepackage{enumitem}
\newtheorem{axiom}{Axiom}
\newtheorem{lemma}{Lemma}
\newtheorem{theorem}{Theorem}
\newtheorem{definition}{Definition}
\newtheorem{corollary}{Corollary}

\lstdefinelanguage{Jimple}{
  keywords={virtualinvoke,specialinvoke,staticinvoke,interfaceinvoke,
             goto,if,return,new,newarray,newmultiarray,throw,
             null,void,int,boolean,byte,char,short,long,float,double},
  sensitive=true,
  morecomment=[l]{//},
  morestring=[b]"
}

\newcommand{\cmb}{\textcolor{black}}

\begin{document}

\title{Increasing the Precision of IR Mapping for Incremental Dataflow Analysis}

\author{Anushri Jana}
\email{anushri.jana@tcs.com}
\affiliation{%
  \institution{TCS Research and DoT SPPU}
  \city{Pune}
  \country{India}
}

\author{Kush Goradiya}
\email{kush.goradiya@tcs.com}
\affiliation{%
  \institution{TCS Research}
  \city{Pune}
  \country{India}
}

\author{Uday P. Khedker}
\email{uday@cse.iitb.ac.in}
\affiliation{%
  \institution{IIT Bombay}
  \city{Mumbai}
  \country{India}}


\begin{abstract}
Static analysis tools operate on an Intermediate Representation~(IR) --- an internal model of the program 
on which dataflow analysis computes and stores facts against IR object identifiers. 
Incremental dataflow analysis reuses  these cached facts across consecutive program versions, but requires an accurate mapping between IR objects in the old and new versions.
When this mapping is imprecise, the consequence is not merely a missed optimization: unmapped IR objects inflate
the \emph{change set}, increasing the number of program points 
that the analyzer must recompute and forcing it to redo work that a more precise mapping would have avoided.
This cost is not specific to any one kind of IR object; it arises wherever IR-level identity is unstable across versions.

\cmb{Any IR that decomposes nested source expressions generates temporary variables which }
 are assigned names by a method-local counter that shifts whenever
a statement is inserted or deleted earlier in the method, breaking any name- or attribute-based
mapping. 
We present a systematic algorithm that establishes a
sound, one-to-one mapping of these temporary variables across consecutive versions, and
 formalize and prove the soundness of the mapping,
guaranteeing that every produced mapping is semantically correct.
The resulting precise mapping shrinks the change set, reducing the number of program points the
incremental analysis must recompute. 
Implemented within the Soot framework \cmb{ for Jimple IR} and evaluated on 8 real-world
 Java projects with genuine edits from the BUGSWARM dataset,
our approach reduces the IR change set by over~82\% on average compared
to Reviser's graph-differencing algorithm, while
completing the full pipeline within~250\,ms per project---directly countering the common 
assumption that a more precise change set necessarily costs more time to compute.

\smallskip\noindent\textbf{Keywords:} incremental dataflow analysis,
IR mapping, temporary variables, Jimple IR.
\end{abstract}

\maketitle





%

\section{Introduction}
\label{sec:intro}

Modern software systems evolve continuously. After every commit, analysis tools
must re-examine the codebase to detect bugs, security vulnerabilities, or
correctness violations. Performing a reanalysis from scratch after each change
is prohibitively expensive for large codebases with frequent code changes.
\emph{Incremental dataflow analysis} avoids this cost  by identifying which
program points must be reset and recomputed and reusing cached results at all
others, rather than reanalyzing the entire codebase
~\cite{Arzt2014Reviser,jana2025common,szabo2021incremental,nougrahiya2025incidfa,stein2021demanded,van2020incremental,wang2025incremental}.
The standard approach is to identify the \emph{change set}---the program points
found to mismatch between two program representations---and derive a broader
\emph{affected region}~\cite{jana2025common} by taking, typically, all program points
reachable from the change set via the analysis dependency structure. Affected
region nodes are reset and recomputed; all others reuse cached results. Since
the affected region is derived from the change set, an imprecise (inflated)
change set necessarily yields a correspondingly larger affected region.

\paragraph{IR and dataflow analysis.}
Static analysis tools operate on an \emph{Intermediate Representation}
(IR)---an internal, language-independent model of a program that exposes
its structure and dataflow in a form amenable to analysis.
Soot~\cite{vallee2010soot} is a widely used Java bytecode analysis
framework whose primary IR, Jimple, translates stack-based JVM bytecode
into a typed, three-address, stack-less linear form. To achieve this
simplification, Jimple decomposes complex, nested Java expressions into
sequences of simple statements, each performing at most one operation.
\emph{Temporary variables}---automatically generated locals prefixed
with \texttt{\$} (e.g., \texttt{\$i0}, \texttt{\$stack1})---hold
intermediate results and references for method invocations. These
temporaries are purely an artefact of the IR lowering process; unlike
source-named locals whose identifiers are stable across versions, they
have no counterpart in the source code and their names are assigned by a
method-local counter that increments with each new temporary created
within a method.

Dataflow analysis computes, for each program point, an abstract description of
the program state---such as which variables may be null or which expressions
are constant. In Jimple, each program point corresponds directly to a Jimple
statement, and analysis facts are expressed in terms of the variables visible
at that point---including temporary variables. For example, a nullness analysis
may record that \texttt{\$stack7} is non-null at a given program point, or a
constant propagation analysis may record that \texttt{\$stack3} holds the
value~42. These facts are stored and propagated using variable names as
symbolic keys. As a result, when a temporary is renamed between versions---say
\texttt{\$stack7} becomes \texttt{\$stack8}---the cached fact associated with
the old name cannot be automatically transferred to the new version, even if
both temporaries hold the result of the same computation. The consequence is
not merely a missed reuse opportunity: an unmapped IR object enters the
\emph{change set}---the set of program points directly found to mismatch
between versions. Since cached facts elsewhere may depend on these points, an
incremental analyzer derives a broader \emph{affected
region}~\cite{jana2025common} from the change set and resets and recomputes every
node in that region.  An inflated change set therefore does not merely add a
few extra nodes: it cascades into a correspondingly inflated affected region,
compounding the cost of imprecise mapping.  This cost is not specific to any
one kind of IR object---it arises wherever IR-level identity is unstable across
versions, and the precision of IR mapping therefore has a direct, measurable
effect on how much an incremental analyzer must recompute.

\paragraph{The mapping prerequisite.}
The key enabler of incremental dataflow analysis is an accurate mapping between
IR objects in the old and new versions of a program, so that analysis facts
indexed against old identifiers can be transferred to their counterparts in the
new version. This requirement has been explicitly identified and addressed in
prior work. Ghime et~al.~\cite{ghime2022ir} formalised it as \emph{IR Mapping}
for a proprietary IR targeting COBOL programs and proved that without a correct
mapping, cached results cannot be safely transferred. Nichols
et~al.~\cite{nichols2019fixpoint} addressed the same requirement for
incremental JavaScript analysis through heuristic matching of IR entities
across versions. In both cases, the mapping problem is tractable because the
entities being matched carry stable, source-derived identities. The present
paper addresses this same requirement in the Java/Jimple setting, where the
generated temporaries have no stable identity and name-based or attribute-based
matching is inapplicable.

\cmb{Any IR that decomposes nested source expressions into simple statements
must hold the intermediate results somewhere, so such IRs generate temporary
variables: Jimple creates \texttt{\$}-prefixed locals, LLVM creates unnamed
\texttt{\%$n$} registers, and SSA construction in frameworks such as WALA
creates value numbers rendered as \texttt{v1}, \texttt{v2}, and so on. In each
case a counter assigns the names by position, so an inserted statement
reindexes every temporary that follows it, and name-based matching fails for
all of them. 
We study the problem in the Soot/Jimple setting.}
Table~\ref{tab:motivating} illustrates the core difficulty. When a new
statement is inserted before \texttt{x = x + y + z} (Version~1), the Jimple IR
for Version~2 reindexes the temporary from \texttt{\$stack7} to
\texttt{\$stack8}. Although the sub-expression \texttt{x+y} is semantically
identical in both versions, string-based comparison treats the two IR
statements as entirely different, incorrectly marking line-8 as changed even
though the source is unchanged. 

\begin{table}[t]
\centering
\caption{Java code and corresponding Jimple IR for two versions.
Temporary variables in \textcolor{red}{red} are mapped by our
algorithm despite renaming. Source-line tags (shown as leading numbers)
facilitate alignment between IR statements and originating Java lines.}
\label{tab:motivating}
\scriptsize
\begin{tabular}{@{}p{0.46\columnwidth}p{0.46\columnwidth}@{}}
\toprule
\textbf{Java Version 1} & \textbf{Java Version 2 (insertion)} \\
\midrule
\begin{lstlisting}[language=Java,numbers=none,frame=none,
                   basicstyle=\ttfamily\scriptsize]
...
7: x = x + y + z;
...
\end{lstlisting}
&
\begin{lstlisting}[language=Java,numbers=none,frame=none,
                   basicstyle=\ttfamily\scriptsize]
...
7: System.out.println(y);
8: x = x + y + z;
...
\end{lstlisting}
\\
\midrule
\textbf{Jimple IR Version 1} & \textbf{Jimple IR Version 2} \\
\midrule
\begin{lstlisting}[language=Jimple,numbers=none,frame=none,
                   basicstyle=\ttfamily\scriptsize]
7: $stack7 = x + y;
7: x = $stack7 + z;
\end{lstlisting}
&
\begin{lstlisting}[language=Jimple,numbers=none,frame=none,
                   basicstyle=\ttfamily\scriptsize]
7: $stack7 = <System: ...>;
7: virtualinvoke ...
8: $stack8 = x + y;
8: x = $stack8 + z;
\end{lstlisting}
\\
\bottomrule
\end{tabular}
\hfill \\
\noindent\textit{Our algorithm maps}
\texttt{\textcolor{red}{\$stack7}} (v1)
$\mapsto$ \texttt{\textcolor{red}{\$stack8}} (v2),
correctly identifying line 8 as unchanged.
\end{table}

\paragraph{Consequence for incremental analysis.}
Since generated temporary names are assigned positionally, a single statement
insertion causes the method-local counter to shift, renaming every temporary
introduced after the insertion point. Any technique that identifies changed
code by string-matching IR statements classifies every renamed temporary as a
new, unmatched statement---even though the underlying computation is
identical---and since the renaming cascades through all subsequent statements
that use the renamed temporary as an operand, the resulting change set inflates
well beyond the genuine change.  Reviser's graph differencing
algorithm~\cite{Arzt2014Reviser, Arzt2013ReviserTR} is one such technique: in
our experiments, it reported up to 219 changed source lines for a program where
only 13 lines were genuinely modified---a factor of $17\times$ inflation, found
by direct string-comparison alone, before any reachability-based propagation
(Section~\ref{sec:relwork} discusses Reviser's graph differencing algorithm and
design philosophy in detail). Since Reviser derives its affected region from
this change set via reachability, the inflated change set produces a
correspondingly larger---likely larger still---affected region, so the
$17\times$ figure is a lower bound on the true cost in wasted recomputation.
Every spurious entry forces recomputation of analysis facts for statements
whose underlying logic is unmodified, negating the efficiency benefit of
incremental analysis. 
\cmb{Recomputation cost is not the only consequence. An incremental analysis
stores its facts against IR identifiers, so when the mapping fails to relate a
renamed temporary to its counterpart, the facts keyed on it do not carry over
and the analysis reports them as new/updated. In companion work on a pointer
analysis we observed exactly this: an unrelated edit renamed temporaries, and
the analysis reported changed facts in parts of the program the edit never
affects. An inflated change set costs the incremental efficiency; inflated
changed facts cost it credibility, since the report attributes findings to the
code the commit never affected and offers the reviewer no way to tell these
apart from the ones it did.  }
 Our results show that the common assumption---that a
more precise change set necessarily costs more time to compute---does not hold:
our approach identifies a strictly more precise change set while completing in
significantly less time.

For the purpose of comparison in this paper, we isolate and re-implement
Reviser's graph differencing component~\cite{Arzt2013ReviserTR}, which is
responsible for change set computation and is the part directly affected by
temporary variable reindexing. We do not use Reviser's incremental IFDS/IDE
solver or any other component of the Reviser system.

\paragraph{Our approach.}
Our key insight is that Soot embeds source-line tags in the IR, recording
which source line each Jimple statement was generated from. For source
lines confirmed as textually unchanged by a source-level diff, Soot
deterministically generates the same sequence of Jimple statements---in
the same order and of the same types. Within this sequence, temporary
variables first appear only on the LHS of assignment statements; their names
differ between versions due to reindexing, but their RHS expressions are
semantically equivalent. We say semantically rather than textually
equivalent because some RHS expressions use a previously introduced
temporary as an operand---and that temporary will have been renamed.
However, since our algorithm processes the statement sequence in order
and builds the mapping incrementally, any temporary appearing in a RHS
will already have a mapping entry established from an earlier statement
in the same sequence. Under this mapping, the RHS are confirmed
equivalent, and the LHS temporaries are therefore correctly mapped.
This gives the approach its soundness: every produced mapping pair
corresponds to the same computed value at the same logical program point.

\paragraph{Contributions.}
\begin{enumerate}
  \item A source-anchored temporary variable mapping algorithm for Jimple IR
that establishes one-to-one mapping between temporaries in consecutive program
versions, directly reducing the unnecessary recomputation that an imprecise
change set would otherwise force on an incremental analyzer
(Section~\ref{sec:approach}).

  \item A soundness proof of the mapping, formalised via three axioms, three
supporting lemmas, and a main theorem, guaranteeing that every produced mapping
pair is semantically correct (Section~\ref{sec:soundness}).

  \item An empirical evaluation on 8 real-world Java projects with genuine
edits from the BUGSWARM dataset, demonstrating our algorithm reduces IR change
set by over 82\% on average compared to Reviser's graph differencing algorithm
(recall\, $=1.0$ in all cases), while completing the full pipeline---source
diffing, edited method identification, and temporary variable mapping---within
under 250\,ms per project, directly countering the assumption that a more
precise change set requires more time than a conservative one
(Section~\ref{sec:eval}).

\end{enumerate}

\paragraph{Paper organisation.}
Section~\ref{sec:relwork} reviews related work.
Section~\ref{sec:approach} presents the problem statement and algorithm.
Section~\ref{sec:soundness} proves soundness.
Section~\ref{sec:eval} reports experiments.
Section~\ref{sec:conclusion} concludes.


\section{Related Work}
\label{sec:relwork}

\subsection{Incremental Program Analysis}
The general problem of imprecise IR-level identity motivates a range of
incremental analysis strategies, which differ in how (or whether) they derive
a precise change set.  Several tools take two program versions and derive the change
set by comparing their IR or AST representations.
Reviser~\cite{Arzt2014Reviser} applies its graph differencing
algorithm~\cite{Arzt2013ReviserTR} to the per-method control flow graphs (CFGs)
of both versions, matching nodes via string-based comparison of their Jimple
statement representations; nodes whose statement strings do not match form the
change set.  Reviser's affected region is then derived by taking all nodes
transitively reachable from the change set via the CFG successor relation, and
affected region nodes are reset and scheduled for recomputation.  Reviser
follows a clear-and-propagate philosophy~\cite{Arzt2014Reviser} that
deliberately accepts an over-approximated change set in exchange for a simple,
string-based differencing step---under the implicit assumption that identifying
a precise change set would require more computation than a conservative one.
For the purpose of comparison in this paper, we isolate and re-implement Reviser's graph
differencing component~\cite{Arzt2013ReviserTR}, which is responsible for
change set computation and is the part directly affected by temporary variable
reindexing.  When a source-level edit causes Jimple to reindex temporaries,
string-based statement matching treats every reindexed statement as mismatched,
inflating the change set itself---before any reachability step is even applied.
Since Reviser's affected region is derived from this change set by
reachability, the inflation compounds further downstream. Our experimental
results directly counter Reviser's design assumption: our mapping-based
approach produces a strictly more precise change set than Reviser's graph
differencing algorithm while completing in significantly less time on all 8
evaluated projects (Section~\ref{sec:eval}).

\emph{truediff}~\cite{erdweg2021concise} similarly takes two source versions,
computes an edit script of added and removed AST nodes as the change set, and
passes this to the IncA incremental Datalog
framework~\cite{szabo2021incremental} for affected region derivation and
recomputation.

Some incremental analysis tools sidestep change set identification entirely
because the change is available by other means. Tools using change
expressions~\cite{van2020incremental} or operating on mutable program
representations---such as Demanded Abstract
Interpretation~\cite{stein2021demanded}, which edits a live Demanded Abstract
Interpretation Graph (DAIG) directly---have the change implicit in the edit
itself. Classical incremental analysis works such as Ryder and
Paull~\cite{ryder1988incremental} and Pollock and
Soffa~\cite{pollock1989incremental}, as well as
IncIDFA~\cite{nougrahiya2025incidfa}, assume change set is supplied externally
by an upstream component. For all these tools, deriving change set from two
program versions is outside their stated scope.

\subsection{IR Mapping}
\label{sec:relwork:irmapping}

Ghime et~al.~\cite{ghime2022ir} formalised IR mapping as a named prerequisite
for incremental static analysis, working with a proprietary IR for COBOL
programs. In their setting, every IR object corresponds to a named source
entity---a variable, a function, an expression---and retains that name across
program versions. Their matching algorithm exploits this stability: for a
textually unchanged source line, IR objects are matched by name, type, and
order of appearance, and their correctness proof rests on the fact that these
attributes are stable for unchanged entities.

This approach is not applicable to the Java/Jimple setting for a fundamental
reason: Jimple generated temporaries carry no source-derived name. A temporary
\texttt{\$stack7} in $v_1$ becomes \texttt{\$stack8} in $v_2$ for the same
computation---not because the program changed, but because the method-local
counter shifted due to an edit elsewhere in the method. Attribute-based
matching would find no name match and produce no mapping, or worse, produce
incorrect mappings by coincidental counter overlap. The instability that
Ghime et~al.\ sidestep entirely is precisely the central obstacle in our
setting.

Nichols et~al.~\cite{nichols2019fixpoint} address mapping of IR entities for
incremental JavaScript analysis using edit-distance-based scoring between pairs
of IR entities, with matching declared below an empirically calibrated
threshold. Their approach handles function renaming but cannot guarantee
correctness of the produced mappings, as an incorrect match causes cached facts
from a semantically different program point to be transferred to the wrong
location in the new version.

The present paper addresses IR mapping in the Java/Jimple setting, where
neither attribute-based matching nor threshold-based heuristic matching is
applicable. Our solution anchors mapping in Soot's source-line tags and
Jimple's deterministic IR generation, enabling a sound, threshold-free mapping
of generated temporaries at the individual statement level.

\subsection{Source-Code-Level Differencing}
Myers' algorithm~\cite{myers1986ond}, implemented in java-diff-utils,
efficiently computes minimal edit scripts between sequences of text lines. It
treats code as plain text, ignoring syntactic and semantic structure, and forms
the foundation of our source-level differencing phase.

Token-based tools such as MOSS~\cite{alex1994moss,schleimer2003winnowing}
detect structural similarities via hashed fingerprints of token sequences.
AST-based tools---GumTree~\cite{Falleri2014GumTree}, GumTree
Simple~\cite{falleri2024fine}, HyperAST~\cite{le2022hyperast},
HyperDiff~\cite{le2023hyperdiff}, and
ChangeDistiller~\cite{Fluri2007ChangeDistilling}---provide fine-grained
source-level change detection by comparing syntactic tree structures. While
these tools produce precise source-level diffs, they identify changes at the
source construct level and do not establish IR-level object correspondences.
They therefore cannot directly support reuse of dataflow facts computed on IR
entities such as Jimple temporaries.

\subsection{Bytecode-Level Differencing}

SootDiff~\cite{Dann2019SootDiff} and jNorm~\cite{schott2024java} address
bytecode similarity and equivalence analysis---comparing two bytecode artefacts
to determine whether they are functionally equivalent, for tasks such as code
clone detection or compiler output verification. SootDiff demonstrated that
differencing at the Jimple IR level is more robust than raw JVM bytecode, using
Jimple as a normalisation layer. jNorm achieves compiler-agnostic bytecode
normalisation, handling discrepancies introduced by different Java compilers.
Both tools apply Myers' algorithm to sequences of IR statements and therefore
encounter the same string-based matching fragility: temporary variable
reindexing causes entire method suffixes to be classified as changed. Neither
tool is designed for incremental analysis and neither provides a variable-level
mapping between versions.

\section{Approach}
\label{sec:approach}

\subsection{Problem Statement}
\label{sec:approach:problem}

\begin{definition}[Temporary Variable Mapping]
Given two consecutive versions $v_1$ and $v_2$ of a Java program and
their corresponding Jimple IRs $\mathit{IR}_1$ and $\mathit{IR}_2$,
let $\mathcal{T}_1$ and $\mathcal{T}_2$ denote the sets of temporary
variables in $\mathit{IR}_1$ and $\mathit{IR}_2$ respectively.
A \emph{temporary variable mapping} is a partial injection:
\[
  \mathsf{tmpVarMap} \;:\; \mathcal{T}_1 \rightharpoonup \mathcal{T}_2
\]
such that $\mathsf{tmpVarMap}(t_1) = t_2$ if and only if $t_1$ and
$t_2$ are \emph{semantically equivalent}---they hold the result of
the same sub-expression at the same logical program point in their
respective versions.
\end{definition}

\noindent A sound mapping enables an incremental analysis framework to
transfer any analysis fact $f$ associated with $t_1$ in $v_1$ directly
to $\mathsf{tmpVarMap}(t_1)$ in $v_2$ without recomputation.

We require the mapping to be sound: every pair $(t_1,t_2)$ in
$\mathsf{tmpVarMap}$ must be semantically equivalent. An unsound mapping would
cause the incremental analyzer to silently transfer dataflow facts computed for
$t_1$ to a semantically different temporary $t_2$, propagating incorrect
abstract states to all dependent program points.  The incremental analysis would
then produce results inconsistent with a from-scratch analysis of
$v_2$, violating the fundamental correctness requirement of incremental
analysis: \emph{from-scratch consistency}~\cite{jana2025common}. 

\subsection{Overview}
\label{sec:approach:overview}

Our methodology proceeds in two phases:

\paragraph{Phase~1---Source-Level Differencing.}
We apply a source-level diff utility
(java-diff-utils\footnote{\url{https://github.com/java-diff-utils/java-diff-utils}},
a Java implementation of Myers' algorithm~\cite{myers1986ond}) to the old and
new source files. This produces two outputs: (i)~a \emph{line-number map},
denoted as $\mathsf{lineNoMap}$, associating each unchanged line number in
$v_2$ with its corresponding line number in $v_1$, and (ii)~two sets of changed
line numbers --- \emph{deletion list}, denoted as $\mathit{delList}$,
comprising line numbers present in $v_1$ but absent in $v_2$, and an
\emph{insertion list}, denoted as $\mathit{insrtList}$, comprising of line
numbers present in $v_2$ but absent in $v_1$ . Modifications are treated as a
deletion followed by an insertion.

\paragraph{Phase~2---Temporary Variable Mapping.}
We recompile the edited source to obtain the updated Jimple IR and identify the
edited method(s). Using $\mathsf{lineNoMap}$ as alignment anchors, we traverse
the IR statements of the edited method in parallel, aligning IR statements by
their source-line tags. For IR statements confirmed as originating from
unchanged source lines---as determined by $\mathsf{lineNoMap}$--- we attempt to
map their LHS temporary variables. IR statements originating from lines present
in $\mathit{delList}$, $\mathit{insrtList}$ or both are skipped without
mapping.

\subsection{Algorithm}
\label{sec:approach:algorithm}

Algorithm~\ref{alg:mapping} formalises Phase~2.

Algorithm~\ref{alg:mapping} proceeds as follows. It begins by invoking
\textsc{ComputeLineMapping} on the source files $v_1$ and $v_2$, producing
$\mathsf{lineNoMap}$.  It then invokes \textsc{ComputeChanges} to classify
changed line numbers into $\mathit{delList}$ and $\mathit{insrtList}$, and
\textsc{FetchEditedMethod} to identify the edited method using these two lists.
The IR statements of the edited method are retrieved from ${IR}_1$ and ${IR}_2$
into $\mathit{stmts_1}$  and $\mathit{stmts_2}$ respectively, and two indices
$i$ and $j$ are initialised to traverse these lists in parallel.

In each iteration of the while loop, the algorithm fetches the current IR
statements $s_1$ and $s_2$ at positions $i$ and $j$, and retrieves their
associated source line numbers $l_1$ and $l_2$ via \textsc{srcLineNo}. The core
logic is a four-way branch:

\begin{itemize}

\item Unchanged line — if $\mathsf{lineNoMap}[\ell_1] = \ell_2$, the two
statements originate from textually identical source lines.
\textsc{MapTempVars} is invoked on ($s_1$, $s_2$), and both $i$ and $j$ are
advanced together.

\item Modified line — if $\ell_1 = \ell_2$ and $\ell_1 \in
\mathit{delList}$ and $\ell_2 \in \mathit{insrtList}$, the line was
modified between versions. All IR statements associated with that line number —
including any statements carrying tag -1, denoting compiler-generated
statements with no direct source-line correspondence--- are skipped on both
sides by advancing $i$ and $j$ independently.

\item Deleted line — if $\ell_1 \in \mathit{delList}$ only, the line
exists in $v_1$ but not in $v_2$. Index $i$ is advanced to bypass all IR
statements associated with that line on the old side.

\item Inserted line — if $\ell_1 \in \mathit{insrtList}$ only, the
line exists in $v_2$ but not in $v_1$. Index $j$ is advanced to bypass all IR
statements associated with that line on the new side.

\end{itemize}

Within \textsc{MapTempVars}, the algorithm checks whether both $s_1$ and $s_2$
are assignment statements (\texttt{AssignStmt}). If so, it extracts their LHS operands into
$\mathit{lhs_1}$ and $\mathit{lhs_2}$. If both are confirmed as temporary variables (variable
name starting with \$) by \textsc{IsTempVar}, a mapping entry
$\mathsf{tmpVarMap}[\mathit{lhs}_1] \leftarrow \mathit{lhs}_2$ is recorded.

\begin{algorithm}[t]
\caption{Temporary variable mapping via source-line anchors}
\label{alg:mapping}
\begin{algorithmic}[1]
\REQUIRE Source files $v_1$, $v_2$; Jimple IRs $\mathit{IR}_1$, $\mathit{IR}_2$
\ENSURE $\mathsf{tmpVarMap}$: mapping of temporaries between versions
\STATE $\mathsf{lineNoMap} \leftarrow \textsc{ComputeLineMapping}(v_1, v_2)$
\STATE $(\mathit{delList}, \mathit{insrtList}) \leftarrow \textsc{ComputeChanges}(v_1, v_2)$
\STATE $\mathit{mthd} \leftarrow \textsc{FetchEditedMethod}(\mathit{delList}, \mathit{insrtList})$
\STATE $\mathit{stmts}_1 \leftarrow \mathit{IR}_1.\textsc{getIRStmt}(\mathit{mthd})$
\STATE $\mathit{stmts}_2 \leftarrow \mathit{IR}_2.\textsc{getIRStmt}(\mathit{mthd})$
\STATE $i \leftarrow 0$; $j \leftarrow 0$
\WHILE{$i < |\mathit{stmts}_1|$ \textbf{and} $j < |\mathit{stmts}_2|$}
  \STATE $s_1 \leftarrow \mathit{stmts}_1[i]$; $s_2 \leftarrow \mathit{stmts}_2[j]$
  \STATE $\ell_1 \leftarrow s_1.\textsc{srcLineNo}()$; $\ell_2 \leftarrow s_2.\textsc{srcLineNo}()$
  \IF{$\mathsf{lineNoMap}[\ell_1] = \ell_2$}
    \STATE \COMMENT{Unchanged line: map temporaries}
    \STATE $\textsc{MapTempVars}(s_1, s_2)$
    \STATE $i \leftarrow i+1$; $j \leftarrow j+1$
  \ELSIF{$\ell_1 = \ell_2$ \textbf{and} $\ell_1 \in \mathit{delList}$ \textbf{and} $\ell_2 \in \mathit{insrtList}$}
    \STATE \COMMENT{Modified line: skip both sides}
    \WHILE{$i < |\mathit{stmts}_1|$ \textbf{and} ($\mathit{stmts}_1[i].\textsc{srcLineNo}() = \ell_1$ \textbf{or} $\mathit{stmts}_1[i].\textsc{srcLineNo}() = -1$)}
      \STATE $i \leftarrow i+1$
    \ENDWHILE
    \WHILE{$j < |\mathit{stmts}_2|$ \textbf{and} ($\mathit{stmts}_2[j].\textsc{srcLineNo}() = \ell_2$ \textbf{or} $\mathit{stmts}_2[j].\textsc{srcLineNo}() = -1$)}
      \STATE $j \leftarrow j+1$
    \ENDWHILE
  \ELSIF{$\ell_1 \in \mathit{delList}$}
    \STATE \COMMENT{Deleted line: advance old side}
    \STATE $i \leftarrow i+1$
  \ELSIF{$\ell_2 \in \mathit{insrtList}$}
    \STATE \COMMENT{Inserted line: advance new side}
    \STATE $j \leftarrow j+1$
  \ENDIF
\ENDWHILE
\medskip
\STATE \textbf{procedure} \textsc{MapTempVars}$(s_1, s_2)$:
\IF{$s_1$ \textbf{is} \texttt{AssignStmt} \textbf{and} $s_2$ \textbf{is} \texttt{AssignStmt}}
  \STATE $\mathit{lhs}_1 \leftarrow s_1.\textsc{getLeftOp}().\textsc{toString}()$
  \STATE $\mathit{lhs}_2 \leftarrow s_2.\textsc{getLeftOp}().\textsc{toString}()$
  \IF{$\textsc{IsTempVar}(\mathit{lhs}_1)$ \textbf{and} $\textsc{IsTempVar}(\mathit{lhs}_2)$}
    \STATE $\mathsf{tmpVarMap}[\mathit{lhs}_1] \leftarrow \mathit{lhs}_2$
  \ENDIF
\ENDIF
\end{algorithmic}
\end{algorithm}

\paragraph{Handling of language constructs.}
\cmb{The algorithm requires no per-construct treatment. Since it anchors
alignment on source-line tags rather than on control-flow structure, the same
traversal handles branches, loops, and exception handlers: their bodies carry
line tags like any other statement, so it maps them when their source lines are
unchanged and skips them otherwise. Lambda expressions and anonymous inner
classes become separate methods and classes in bytecode, each a method scope to
which Algorithm~\ref{alg:mapping} applies unchanged. One caveat: a counter
assigns the names of these synthetic methods and classes by position too, so
adding a lambda shifts the index of those that follow, and the instability we
resolve for temporary variables recurs when identifying the edited scope.  We
do not currently map synthetic method identity across versions; extending the
source-anchored approach to it is future-work.  }

\paragraph{Correctness Guarantee of \textsc{MapTempVars} Invocation.}
\textsc{MapTempVars} is invoked exclusively from the branch guarded by the
condition $\mathsf{lineNoMap}[\ell_1] = \ell_2$ (line~10 of
Algorithm~\ref{alg:mapping}). This condition is satisfied only when $l_1$ and
$l_2$ are confirmed as textually identical by java-diff-utils, as recorded in
$\mathsf{lineNoMap}$. All remaining branches — the modified-line branch, which
advances past statements whose line numbers appear in both $\mathit{delList}$ and
$\mathit{insrtList}$; the deleted-line branch, which advance past statements whose
line numbers appear only in $\mathit{delList}$; and the inserted-line branch,
which advances past statements whose line numbers appear only in
$\mathit{insrtList}$ — advance the indices $i$ and $j$ without invoking
\textsc{MapTempVars} and without recording any entry in $\mathsf{tmpVarMap}$.
Consequently, every pair ($t_1$, $t_2$)  recorded in $\mathsf{tmpVarMap}$
originates exclusively from IR statements confirmed as unchanged, which is the basis
of the soundness argument in Section~\ref{sec:soundness}.

\paragraph{Complexity.} The algorithm traverses $\mathit{stmts}_1$ and
$\mathit{stmts}_2$ exactly once, giving a time complexity of
$O(|\mathit{stmts}_1| + |\mathit{stmts}_2|)$---linear in the size of the edited
method's IR. This bound holds because each of the four branches advances at
least one of $i$ or $j$, ensuring the loop terminates and no statement is processed
more than once.  The source-level diff (Phase~1) runs in $O(NM)$ worst case via
Myers' algorithm, but in practice completes in milliseconds for typical Java
files.

\paragraph{Output Use.}
The resulting $\mathsf{tmpVarMap}$ is the direct input to an
incremental analysis framework: for any dataflow fact $f$ associated
with temporary $t_1$ in $v_1$, the framework transfers $f$ to
$\mathsf{tmpVarMap}(t_1)$ in $v_2$ without recomputation.

\section{Soundness of the Mapping}
\label{sec:soundness}

We now prove that every pair $(t_1, t_2)$ produced by
Algorithm~\ref{alg:mapping} is a correct mapping---i.e.,
$t_1$ and $t_2$ hold the result of the same sub-expression at
the same logical program point.

\subsection{Axioms}

\begin{axiom}[Source-diff correctness]
\label{ax:diff}
java-diff-utils correctly identifies unchanged source lines.
A line $\ell_2$ in $v_2$ is associated with $\ell_1$ in $v_1$ via
$\mathsf{lineNoMap}[\ell_1] = \ell_2$ if and only if the textual
content of line~$\ell_1$ in $v_1$ equals that of line~$\ell_2$
in~$v_2$ (modulo cosmetic whitespace).
\end{axiom}

\begin{axiom}[Soot line-tag faithfulness]
\label{ax:tags}
Soot's source-line tags faithfully record provenance.
If a Jimple statement $s$ carries tag $\ell$, then $s$ was generated
from source line~$\ell$.
\end{axiom}

\begin{axiom}[Deterministic IR generation for unchanged lines]
\label{ax:det}
For any source line whose textual content is identical in $v_1$ and $v_2$, Soot
generates the same sequence of Jimple statements --- same count, same statement
types, in the same order. Within this sequence, each temporary variable is
\emph{introduced} exactly once on the LHS of an \texttt{AssignStmt}, and may
subsequently appear as an operand on RHS of later \texttt{AssignStmt}s in the
same sequence.  All non-temporary operands (source-named variables, constants,
method signatures) in every statement are textually identical across versions.
The only variation between versions is in the \emph{names} of the temporary
variables, which are assigned by a method-local counter that may shift due to
edits elsewhere within the method.
\end{axiom}

\noindent Axiom~\ref{ax:diff} is a property of java-diff-utils.
Axiom~\ref{ax:tags} is a property of Soot's front-end.  Axiom~\ref{ax:det}
follows from the determinism of the compilation pipeline: identical source
text compiled with the same compiler and settings produces identical JVM
bytecode for the corresponding method, and Soot's Jimple generator
deterministically lifts identical bytecode into the same sequence of Jimple
statements.  The only variation arises from the method-local temporary index
counter, which is method-scoped and may shift due to edits elsewhere in the
method body, producing different temporary names while leaving all other
aspects of the generated statements unchanged.

\subsection{Lemmas}

\begin{lemma}[Structural identity of unchanged lines]
\label{lem:structural}
Let $\ell_1 \in v_1$ and $\ell_2 \in v_2$ satisfy
$\mathsf{lineNoMap}[\ell_1] = \ell_2$.
Let $\mathcal{S}_{\ell_1}$ and $\mathcal{S}_{\ell_2}$ denote the
sequences of Jimple IR statements generated from $\ell_1$ in $\mathit{IR}_1$
and from $\ell_2$ in $\mathit{IR}_2$ respectively.
Then:
\begin{description}
  \item[(i)] $|\mathcal{S}_{\ell_1}| = |\mathcal{S}_{\ell_2}|$ ---
        the sequences have equal length;
  \item[(ii)] for each position $k$, the $k$-th statement in
        $\mathcal{S}_{\ell_1}$ and $\mathcal{S}_{\ell_2}$ have
        the same statement type;
  \item[(iii)] for each \texttt{AssignStmt} at position $k$, the LHS
        operands differ only in temporary variable names; and
  \item[(iv)] for each \texttt{AssignStmt} at position $k$, the RHS
        expression is semantically equivalent under the
        partial mapping $\mathsf{tmpVarMap}$ constructed for
        positions $1, \ldots, k{-}1$ on $\ell_1$ and $\ell_2$:
        all non-temporary operands in the RHS are textually
        identical across versions, and any temporary operand
        $t \in \mathcal{T}_1$ appearing in the RHS satisfies
        $\mathsf{tmpVarMap}[t] = t'$, where $t'$ is the
        corresponding operand in $\mathit{IR}_2$.
\end{description}
\end{lemma}

\begin{proof}
By Axiom~\ref{ax:diff}, lines $\ell_1$ and $\ell_2$ have identical
textual content.
By Axiom~\ref{ax:tags}, all IR statements tagged $\ell_1$
(resp.~$\ell_2$) were generated from that common source text.
By Axiom~\ref{ax:det}, Soot deterministically generates the same
structural IR from identical source text.

\noindent\textit{Claims~(i) and~(ii)} follow directly from
Axiom~\ref{ax:det}: identical source text produces the same number
of statements in the same order with the same types.

\noindent\textit{Claim~(iii)} follows from Axiom~\ref{ax:det}:
the only variation on the LHS is the method-local temporary index
counter, which shifts due to edits elsewhere in the method but
does not affect statement structure.

\noindent\textit{Claim~(iv)} is established by induction on
position $k$ within $\mathcal{S}_{\ell_1}$ and
$\mathcal{S}_{\ell_2}$.

\smallskip\noindent\textit{Base case} ($k = 1$): the RHS of the
first \texttt{AssignStmt} on line $\ell_1$ contains no temporary
introduced earlier on this line. All operands are therefore either
source-named variables, constants, or method signatures, which are
textually identical by Axiom~\ref{ax:det}. Hence the RHS is
semantically equivalent with an empty $\mathsf{tmpVarMap}$.

\smallskip\noindent\textit{Inductive step} ($k > 1$): assume
claim~(iv) holds for all positions $k' < k$. Consider the RHS of
the $k$-th \texttt{AssignStmt}. Any non-temporary operand is
textually identical by Axiom~\ref{ax:det}. Any temporary operand
$t \in \mathcal{T}_1$ in this RHS was introduced on the LHS of
some position $k' < k$ on the same source line $\ell_1$ (since
Jimple temporaries are method-local and introduced by
\texttt{AssignStmt}s). By claim~(iii) applied at $k'$, the
corresponding LHS temporary in $\mathit{IR}_2$ is
$\mathsf{tmpVarMap}[t] = t'$. By the inductive hypothesis at $k'$,
$t$ and $t'$ are semantically equivalent --- they hold the result
of the same sub-expression. Hence replacing $t$ with $t'$ via
$\mathsf{tmpVarMap}$ yields a semantically equivalent RHS in
$\mathit{IR}_2$, establishing claim~(iv) at position $k$.
\end{proof}

\begin{lemma}[Mapping confined to unchanged lines]
\label{lem:confinement}
For every pair $(s_1, s_2)$ for which \textsc{MapTempVars} is invoked
in Algorithm~\ref{alg:mapping}, both $s_1$ and $s_2$ originate from
source lines confirmed as unchanged by $\mathsf{lineNoMap}$.
\end{lemma}

\begin{proof}
Inspect Algorithm~\ref{alg:mapping}. \textsc{MapTempVars} is called
only at line~12, which is reached exclusively from the branch at
lines~10--13, guarded by the condition
$\mathsf{lineNoMap}[\ell_1] = \ell_2$.
By Axiom~\ref{ax:diff}, this condition holds if and only if lines
$\ell_1$ and $\ell_2$ are textually identical.
All other branches---the modified-line branch (lines~14--21),
the deleted-line branch (lines~22--24), and the inserted-line branch
(lines~25--27)---advance the indices $i$ and/or $j$ without invoking
\textsc{MapTempVars}.
\end{proof}

\begin{lemma}[Parallel traversal preserves positional correspondence]
\label{lem:parallel}
When \textsc{MapTempVars}$(s_1, s_2)$ is invoked, $s_1$ and $s_2$
are the $k$-th Jimple statement generated from their respective (mapped)
source lines, for the same index $k$.
\end{lemma}

\begin{proof}
The algorithm advances $i$ and $j$ in lock-step at line~13 whenever
a mapped line is encountered.
The non-mapping branches advance $i$ and $j$ independently to skip
over deleted, inserted, and modified regions.
After all skips, when the condition at line~10 is satisfied again,
both $i$ and $j$ point to the first unprocessed statements from their
respective next mapped line pair.

By Lemma~\ref{lem:structural}, for a mapped line pair $(\ell_1,
\ell_2)$ the two IR sequences have the same length and order.
Since both indices start at the beginning of the respective sequences
for $\ell_1$ and $\ell_2$ and are incremented together in line~13,
the $k$-th invocation of \textsc{MapTempVars} for that line pair
receives the $k$-th statement from each side.
\end{proof}

\subsection{Main Theorem}

\begin{theorem}[Soundness of \textsf{tmpVarMap}]
\label{thm:soundness}
For every mapping $(t_1 \mapsto t_2) \in \mathsf{tmpVarMap}$ produced
by Algorithm~\ref{alg:mapping}, $t_1$ and $t_2$ are semantically
equivalent: they hold the result of the same sub-expression at
corresponding logical program points in $v_1$ and $v_2$ respectively.
\end{theorem}

\begin{proof}
Let $(t_1 \mapsto t_2) \in \mathsf{tmpVarMap}$. This mapping is
produced inside \textsc{MapTempVars}$(s_1, s_2)$ at lines~30--37 of
the procedure, so $t_1 = \mathit{lhs}_1$ and $t_2 = \mathit{lhs}_2$.

\textit{Step 1.} By Lemma~\ref{lem:confinement}, $(s_1, s_2)$ were
passed to \textsc{MapTempVars} only because
$\mathsf{lineNoMap}[\ell_1] = \ell_2$, i.e., lines $\ell_1$ and
$\ell_2$ are textually identical.

\textit{Step 2.} By Lemma~\ref{lem:structural}(ii), $s_1$ and $s_2$
are \texttt{AssignStmt}s of the same type (confirmed also at the
\textsc{IsTempVar} check in \textsc{MapTempVars}).
By Lemma~\ref{lem:structural}(iii), their LHS operands differ only
in temporary variable name, so $t_1$ and $t_2$ are the LHS
temporaries of structurally corresponding assignments.
By Lemma~\ref{lem:structural}(iv), the RHS expression of $s_1$ is
semantically equivalent to the RHS expression of $s_2$ under the
partial $\mathsf{tmpVarMap}$ constructed for all preceding positions
on $\ell_1$ and $\ell_2$: non-temporary operands are textually
identical, and any temporary operand in the RHS of $s_1$ has already
been mapped to its semantically equivalent counterpart in $s_2$ via
$\mathsf{tmpVarMap}$. Hence $s_1$ and $s_2$ compute the same value,
and $t_1$ and $t_2$ hold that value at corresponding program points.

\textit{Step 3.} By Lemma~\ref{lem:parallel}, $s_1$ and $s_2$ are the
same-indexed statements within their respective mapped line sequences,
i.e., they represent the same computational step.

\textit{Conclusion.} $t_1$ holds the result of expression $e$ at
program point $p_1$ in $v_1$; $t_2$ holds the result of the same
expression $e$ at the corresponding program point $p_2$ in $v_2$.
Hence $t_1$ and $t_2$ are semantically equivalent, and the mapping
is sound.

Since this argument applies to every entry in $\mathsf{tmpVarMap}$,
the entire map is sound.
\end{proof}

\begin{corollary}[Safe analysis-fact transfer]
\label{cor:transfer}
Let $f$ be an analysis fact associated with temporary $t_1$ in the
dataflow analysis result for $v_1$. Under the conditions of
Theorem~\ref{thm:soundness}, the incremental analysis framework may
associate $f$ with $\mathsf{tmpVarMap}(t_1)$ in $v_2$ without
recomputation, and this transfer is semantically correct.
\end{corollary}

\begin{proof}
Follows directly from Theorem~\ref{thm:soundness}: since
$\mathsf{tmpVarMap}(t_1)$ computes the same value as $t_1$ at the
corresponding program point, any analysis fact derived from that value
(e.g., non-nullness, constancy, type bounds) applies equally to both.
\end{proof}

\subsection{Scope and Limitations}

The soundness proof is conditional on Axioms~\ref{ax:diff}--\ref{ax:det}.
In practice:
\begin{itemize}
  \item Axiom~\ref{ax:diff} may fail for lines that are textually
        identical but semantically different due to surrounding context
        (e.g., a line \texttt{return x} that appears in two different
        methods but is classified as unchanged). This is mitigated by
        the fact that our mapping is scoped to the \emph{edited method}
        only.
  \item Axiom~\ref{ax:tags} may be imprecise for certain Soot
        transformations that merge or split source lines. In our
        evaluation, we found no such cases, but this remains a
        theoretical limitation.
  \item Axiom~\ref{ax:det} holds for standard Soot compilation.
        Third-party Soot plugins that perform non-deterministic
        transformations could violate it.
  \item Added or deleted methods. When a method is added in
      $v_2$, it has no counterpart in $v_1$; Algorithm~\ref{alg:mapping}
      produces no entries in $\mathsf{tmpVarMap}$ for that method, and
      the incremental analyzer computes its facts from scratch, which
      is safe. When a method is deleted in $v_2$, the cached facts for
      that method in $v_1$ become unreachable and no transfer is
      attempted. A method renaming at the source level is treated as a
      deletion of the old method and an addition of the new one,
      handled conservatively by the same argument. In all three cases
      the algorithm produces no incorrect mappings --- the conservative
      outcome of an empty $\mathsf{tmpVarMap}$ for the affected method
      preserves from-scratch consistency.
      Similarly, a method moved between classes during refactoring is treated
      as a deletion from the source class and an addition in the target class; no
      mapping is attempted and from-scratch consistency is preserved.

\end{itemize}
The algorithm is conservative: in the presence of any doubt (line tags
absent, non-deterministic IR), the relevant statements are left unmapped,
triggering recomputation rather than incorrect reuse.

\section{Evaluation}
\label{sec:eval}

\subsection{Experimental Setup}
\label{sec:eval:setup}

We implemented Algorithm~\ref{alg:mapping} using the
Soot\footnote{\url{https://github.com/soot-oss/soot}} framework for IR
generation. Soot's Jimple IR was used with standard compilation settings; no
Soot plugins or non-standard transformations were applied. For source-level
diffing, we use
java-diff-utils\footnote{\url{https://github.com/java-diff-utils/java-diff-utils}},
a Java implementation of Myers' algorithm~\cite{myers1986ond}.  Each experiment
was allocated a maximum heap size of 2\,GB on a machine with 4 Intel Core
i5-1145G7 cores (8 logical processors), running Windows~11 with OpenJDK~17 and
16\,GB RAM.

We evaluated on 8 real-world open-source Java projects drawn from the
BUGSWARM dataset~\cite{tomassi2019bugswarm}, which provides paired \emph{failed}
(buggy) and \emph{fixed} versions of programs alongside automated build
scripts. Projects range from 285 to 2345 classes
(Table~\ref{tab:dataset}). We selected BUGSWARM because it contains
real-world edits rather than artificial ones, providing a realistic
distribution of change types including single-statement insertions,
condition modifications, and null-check additions~\cite{wang2025incremental}.
\cmb{We confirmed that Soot emits source-line tags for the bodies of branches,
loops, and exception handlers in all subject methods. We excluded version pairs
whose edits add or remove lambda expressions or anonymous inner classes, since
we do not yet map synthetic method identity across versions.
}

For each fail/pass pair, we ran both our mapper and Reviser's graph
differencing algorithm, implemented following the pseudocode
in~\cite{Arzt2013ReviserTR}. The algorithm takes the per-method CFGs of both
old and new program versions as input and traverses them in parallel. Nodes in
the two CFGs are matched using string-based comparison of their Jimple
statement representations: two nodes are considered equivalent if and only if
their statement strings are textually identical. Unmatched nodes are classified
as added or removed, and edges are also noted as added/removed. For our
comparison purpose, we store the line numbers of added/removed nodes as the
change set, measured by direct string-based node matching only, prior to any
reachability-based affected region derivation, to ensure a fair comparison
against our mapper's change set. All reported times exclude bytecode loading
and CFG construction.  Our mapping time includes the entire pipeline:
java-diff-utils applied across the project, edited-method identification, and
the mapping itself.

\begin{table}[h]
\centering
\caption{Dataset characterisation. Edit type, edited method scope, and
temporary variable count in the edited method are reported per project
version pair.}
\label{tab:dataset}
\scriptsize
\begin{tabular}{@{}llrrr@{}}
\toprule
\textbf{Proj.} &
\textbf{\#mthds} & \textbf{Edit type}  & \textbf{Mthd size} & \textbf{\#tmps} \\
 &\textbf{edited} &  & \textbf{(IR statements)} & \textbf{in mthd} \\
\midrule
a~\cite{bugswarm-a} & 1 & Null-check insertion    & 48 & 31   \\
b~\cite{bugswarm-b} & 1 & Statement modification  & 14 & 7     \\
c~\cite{bugswarm-c} & 1 & Condition modification  & 116 & 70   \\
d~\cite{bugswarm-d} & 1 & Null-check insertion   & 113 & 50   \\
e~\cite{bugswarm-e} & 1 & Null-check insertion   & 494 & 366  \\
\multirow{2}{*}{f~\cite{bugswarm-f}} & \multirow{2}{*}{2} & Null-check insertion   & 11 & 4     \\
                                                          & & Statement deletion     & 52 & 27    \\
g~\cite{bugswarm-g} & 1 & Null-check insertion   & 27 & 12    \\
h~\cite{bugswarm-h} & 1 & Null-check insertion   & 282 & 170  \\
\bottomrule
\end{tabular}
\vspace{2pt} \\
\noindent\textit{\#mthds edited}: number of methods containing a
genuine edit; \textit{Mthd size}: total IR statements in the edited method
scope; \textit{\#tmps}: number of temporary variables in the edited
method.
\end{table}

\subsection{Research Questions}

\noindent\textbf{RQ1 (Efficiency):} Is our mapping algorithm fast
enough to be a practical component of an incremental analysis workflow?
We compare runtime against Reviser's graph differencing
algorithm as a baseline.

\noindent\textbf{RQ2 (Change Set Size Reduction):} How much does our
mapping reduce the number of source lines incorrectly reported as
changed, relative to Reviser's graph differencing
algorithm? A smaller change set directly translates
to fewer IR statements requiring reanalysis and more analysis facts being
safely reused without recomputation.

\subsection{Metrics}
\label{sec:eval:metrics}

\begin{itemize}
  \item \textbf{Mapper Time (s):} total wall-clock time for the full
        mapping pipeline (java-diff-utils, edited method identification, and
        temporary variable mapping).

  \item \textbf{Reviser's graph differencing algorithm Time (s):} total wall-clock time for Reviser's
        graph differencing algorithm.

  \item \textbf{Efficiency Gain (\%):}
        \[
          \frac{\text{Reviser's graph differencing algorithm Time} - \text{Mapper Time}}
               {\text{Reviser's graph differencing algorithm Time}} \times 100
        \]

  \item \textbf{Mapper Change Set (M-CS):} number of source lines
        reported as changed by our algorithm.

  \item \textbf{Reviser Change Set (R-CS):} number of source lines
        reported as changed by Reviser's graph differencing
algorithm.

  \item \textbf{Change Set Size Reduction (\%):}
        \[
          \frac{|\text{R-CS}| - |\text{M-CS}|}{|\text{R-CS}|} \times 100
        \]
        Since both our approach and Reviser's graph differencing algorithm traverse all genuinely
        changed lines and neither produces false negatives (verified
        manually as described in
        Section~\ref{sec:eval:verification}), this reduction directly
        corresponds to the elimination of false positives---source lines
        incorrectly reported as changed due to temporary variable
        reindexing. Consequently, recall\,$= 1.0$ for both approaches
        across all projects. We do not report F1 score as a primary
        metric because, with recall fixed at 1.0 for both approaches,
        F1 reduces monotonically to a function of precision alone and
        conveys no additional information beyond change set size
        reduction. We measure change set size directly; we do not separately measure the affected region that an incremental analyzer would derive from it. Since the affected region is derived from the change set (e.g., by reachability), a smaller change set guarantees a smaller affected region, so the reported change set reduction is a conservative lower bound on the true reduction in wasted recomputation.
\end{itemize}

\subsection{Results}
\label{sec:eval:results}

Table~\ref{tab:results} presents per-project results.

\begin{table}[h]
\centering
\caption{Efficiency (RQ1) and change set size reduction (RQ2) results.
Change set size reduction directly corresponds to false positive
elimination given that recall\,$= 1.0$ for both approaches.}
\label{tab:results}
\scriptsize
\begin{tabular}{@{}lrrrrrrrr@{}}
\toprule
\textbf{Proj.} & \textbf{\#cl.} & \cmb{\textbf{\#m.}}
&\textbf{Map.(s)} & \textbf{Rev.(s)} & \textbf{EG(\%)} &
\textbf{M-CS} & \textbf{R-CS} & \textbf{CSR\%} \\
\midrule
a~\cite{bugswarm-a} & 782  & \cmb{30} & 0.241 & 1.243 & 81 &  3 &  52 & 94 \\
b~\cite{bugswarm-b} & 288  & \cmb{2} & 0.210 & 0.397 & 47 &  1 &   2 & 50 \\
c~\cite{bugswarm-c} & 930  & \cmb{24} & 0.259 & 1.864 & 86 &  1 &  36 & 97 \\
d~\cite{bugswarm-d} & 616  & \cmb{40} & 0.211 & 1.088 & 81 &  4 &   7 & 43 \\
e~\cite{bugswarm-e} & 285  & \cmb{199} & 0.155 & 0.882 & 82 & 13 & 219 & 94 \\
f~\cite{bugswarm-f} & 332  & \cmb{25} & 0.244 & 0.700 & 65 &  7 &  65 & 89 \\
g~\cite{bugswarm-g} & 2345 & \cmb{4} & 0.223 & 9.380 & 98 &  2 &  30 & 93 \\
h~\cite{bugswarm-h} & 2326 & \cmb{121} & 0.289 & 6.978 & 96 &  1 &  25 & 96 \\
\midrule
\textbf{Mean} & & & \textbf{0.229} & \textbf{2.816} &
                  \textbf{80} & \textbf{4.0} & \textbf{54.5} & \textbf{82}  \\
\bottomrule
\end{tabular}
\vspace{2pt} \\
\noindent\textit{\#cl.}: number of classes; \cmb{\textit{\#m.}: number of mappings;} \textit{Map.}: mapper time;
\textit{Rev.}: Reviser's graph differencing algorithm time; \textit{EG}: efficiency gain;
\textit{M-CS}/\textit{R-CS}: mapper/Reviser's graph differencing algorithm change set size;
\textit{CSR\%}: change set size reduction.
\end{table}

\paragraph{RQ1---Efficiency.}
Our mean mapping time is 0.229\,s per project, compared to 2.816\,s
for Reviser's graph differencing algorithm---an average efficiency gain of~80\%. Runtime savings reach
98\% for project~g (2345 classes). Critically, our runtime remains
largely \emph{independent of application size}: for projects~g and~h
(each exceeding 2300 classes), our times are 0.223\,s and 0.289\,s
respectively, while Reviser's graph differencing algorithm takes 9.38\,s and 6.978\,s. This
independence arises because Algorithm~\ref{alg:mapping} traverses only the IR statements of
the edited method scope, as identified via $\mathit{delList}$ and
$\mathit{insrtList}$, rather than the entire class hierarchy. While our
largest project contains 2345 classes, the runtime of our approach is
bounded by the size of the edited method scope rather than total
codebase size, suggesting scalability to larger projects; empirical
validation on enterprise-scale codebases remains future work.

\paragraph{RQ2---Change Set Size Reduction.}
Our approach reduces the change set by an average of~82\% (from 54.5 lines for
Reviser's graph differencing algorithm to 4.0 lines for our mapper). Since
recall\,$= 1.0$ for both approaches, this reduction represents direct
elimination of false positives introduced by temporary variable reindexing. In
project~e, Reviser's graph differencing algorithm reports 219 changed lines for
a program with only 13 genuinely modified---a factor-of-17 inflation. Our
approach reports exactly 13.

Our average change set of 4.0 lines versus Reviser's graph differencing
algorithm's 54.5 lines represents a potential reuse improvement of $>$90\% for
the analysis state of unchanged regions. All our generated change sets are
strict subsets of Reviser's graph differencing algorithm's, confirming zero
false negatives. The manual verification protocol is described in
Section~\ref{sec:eval:verification}.

\paragraph{Analysis of Reviser's graph differencing algorithm's over-reporting.}
Sources of over-reporting in Reviser's graph differencing algorithm: (a) temporary variable
reindexing (the primary cause, addressed by our mapping), and (b)
superficial metadata or build-number updates that Reviser's graph differencing algorithm classifies
as edits. Our source-level diff filter eliminates the latter as well,
since metadata changes typically occur in separate files and are not
present in $\mathsf{lineNoMap}$.

\begin{table*}[ht]
\centering
\caption{Representative example from project~c. The only semantic change is the
null-check insertion at line~118.  Reviser's graph differencing
algorithm reports 36 changed lines due to temporary variable reindexing; our
approach reports 1. Classpaths are omitted for readability.}
\label{tab:example}
\scriptsize
\begin{tabular}{@{}p{0.46\textwidth}p{0.46\textwidth}@{}}
\toprule
\textbf{Java Code --- Failed Version} & \textbf{Java Code --- Fixed Version} \\
\midrule
\begin{lstlisting}[numbers=none,frame=none,
  basicstyle=\ttfamily\scriptsize,language=Java]
...
101 return toRuleMatchArray(ruleMatches);
...
117 messages.getString("desc_spelling_short"));
118 if(
  userConfig.getMaxSpellingSuggestions() == 0
  || ruleMatches.size()
  <= userConfig.getMaxSpellingSuggestions())
{
119 List suggestions = getSuggestions(word);
...
\end{lstlisting}
&
\begin{lstlisting}[numbers=none,frame=none,
  basicstyle=\ttfamily\scriptsize,language=Java]
...
101 return toRuleMatchArray(ruleMatches);
...
117 messages.getString("desc_spelling_short"));
118 if(
  userConfig == null ||
  userConfig.getMaxSpellingSuggestions() == 0
  || ruleMatches.size()
  <= userConfig.getMaxSpellingSuggestions())
{
119 List suggestions = getSuggestions(word);
...
\end{lstlisting}
\\
\midrule
\textbf{Jimple IR --- Failed Version} & \textbf{Jimple IR --- Fixed Version} \\
\midrule
\begin{lstlisting}[numbers=none,frame=none,
  basicstyle=\ttfamily\scriptsize,language=Jimple]
...
101 $stack79 = virtualinvoke
      $r3.toRuleMatchArray($stack13)
101 return $stack79
...
116 $stack32 = $r11.messages
117 $stack35 = virtualinvoke
      $stack32.getString("spelling")
117 $stack33 = $r12.messages
117 $stack34 = virtualinvoke
      $stack33.getString("desc_spelling_short")
117 specialinvoke $u1.<init>(
      $r13, sentence, len,
      $stack36, $stack35, $stack34)
118 $stack37 = $r14.userConfig
118 $stack38 = virtualinvoke
      $stack37.getMaxSpellingSuggestions()
118 if $stack38 == 0 goto
118 $stack56 = interfaceinvoke $stack13.size()
118 $stack54 = $r15.userConfig
118 $stack55 = virtualinvoke
      $stack54.getMaxSpellingSuggestions()
118 if $stack56 > $stack55 goto
119 $stack63 = virtualinvoke
      $r16.getSuggestions(word)
...
\end{lstlisting}
&
\begin{lstlisting}[numbers=none,frame=none,
  basicstyle=\ttfamily\scriptsize,language=Jimple]
...
101 $stack80 = virtualinvoke
      $r3.toRuleMatchArray($stack13)
101 return $stack80
...
116 $stack32 = $r11.messages
117 $stack35 = virtualinvoke
      $stack32.getString("spelling")
117 $stack33 = $r12.messages
117 $stack34 = virtualinvoke
      $stack33.getString("desc_spelling_short")
117 specialinvoke $u1.<init>(
      $r13, sentence, len,
      $stack36, $stack35, $stack34)
118 $stack37 = $r14.userConfig
118 if $stack37 == null goto
118 $stack53 = $r15.userConfig
118 $stack54 = virtualinvoke
      $stack53.getMaxSpellingSuggestions()
118 if $stack54 == 0 goto
118 $stack67 = interfaceinvoke $stack13.size()
118 $stack65 = $r16.userConfig
118 $stack66 = virtualinvoke
      $stack65.getMaxSpellingSuggestions()
118 if $stack67 > $stack66 goto
119 $stack57 = virtualinvoke
      $r17.getSuggestions(word)
...
\end{lstlisting}
\\
\midrule
\multicolumn{2}{@{}l@{}}{%
  \textbf{Graph-differencing CS:}
  $\{101,\ldots,118,119,120,\ldots\}$,
  Total size\,$= 36$} \\
\multicolumn{2}{@{}l@{}}{%
  \textbf{Mapper CS:}
  $\{118\}$,
  Total size\,$= 1$} \\
\multicolumn{2}{@{}p{0.92\textwidth}@{}}{%
  \textbf{Mappings:}
  $\{\ldots,\,
  \texttt{\$stack79} \to \texttt{\$stack80},\,
  \texttt{\$stack32} \to \texttt{\$stack32},\,
  \texttt{\$stack33} \to \texttt{\$stack33},\,
  \texttt{\$stack34} \to \texttt{\$stack34},\,
  \texttt{\$stack35} \to \texttt{\$stack35},\,
  \texttt{\$stack63} \to \texttt{\$stack57},\,\ldots\}$} \\
\bottomrule
\end{tabular}
\end{table*}

\subsection{Manual Verification Protocol}
\label{sec:eval:verification}

To verify that no genuine code changes were missed by our approach, two
of the authors independently inspected each fail/pass version pair. For
each pair, the ground-truth change set was established by examining the
Git diff between the two versions at the source level, identifying all
lines with semantic modifications (excluding whitespace, comment, and
metadata-only changes). The mapper's reported change set was then
compared against this ground truth.

In all 8 projects, the mapper's change set was a strict subset of
Reviser's graph differencing algorithm's change set and a superset of the ground-truth semantic
changes, confirming zero false negatives (recall\,$= 1.0$). No
disagreements between the two inspectors were found.
\cmb{We additionally manually verified all the 445 mappings produced by our algorithm across the eight
projects,
confirming in each case that the two temporaries hold the result of the same
sub-expression. 
}

\subsection{Working Example}
\label{sec:eval:example}
Table~\ref{tab:example} shows a representative example from project~c.
The fixed version introduces a single semantic change: a null check for
\texttt{userConfig} at line~118. Although this is the only meaningful
edit, metadata changes elsewhere in the file cause Jimple to reindex
temporary variables throughout the method.

Our approach correctly identifies line~118 as the sole changed line.
For lines~101 and~116--117, which are aligned by $\mathsf{lineNoMap}$,
variable mappings are established: \texttt{\$stack32},
\texttt{\$stack33}, \texttt{\$stack34}, and \texttt{\$stack35}
(line~117) are each mapped to their counterparts with semantically
equivalent RHS expressions in the fixed version. For line~119 (also
aligned), the mapping successfully pairs \texttt{\$stack63} (failed
version) with \texttt{\$stack57} (fixed version) despite the index
shift caused by the inserted null check. 

Reviser's graph differencing algorithm, which lacks this mapping, reports a
change set of 36 lines due to cascading temporary variable reindexing---found
directly by string comparison, before any reachability step.  Since Reviser's
affected region is derived from the change set via CFG reachability, the
inflated change set of 36 lines (rather than genuine 1) directly negates the
efficiency benefit of incremental analysis, with the true cost in wasted
recomputation likely exceeding the 36-line figure once the affected region is
derived.  Our approach not only produces the precise change set of \{118\} but
does so in significantly less time than the graph differencing
algorithm (Table~\ref{tab:results}), countering the general assumption that
computing a precise change set necessarily requires more computation than a
conservative one.

\subsection{Threats to Validity}
\label{sec:eval:threats}

\paragraph{Internal validity.} 
\cmb{The primary internal
validity threat is risk of re-implementation: we re-implement Reviser's graph
differencing algorithm rather than use the original artefact. We attempted to
build the publicly available Reviser implementation but could not obtain a
working build against the current tool-chain versions, so a direct comparison
against the original system was not available to us. We therefore followed the
pseudocode in~\cite{Arzt2013ReviserTR} strictly, with default settings and no
parameter tuning, and restricted the comparison to change-set computation---the
component affected by temporary variables reindexing---rather than the full
incremental solver.  }
\paragraph{External validity.}
\cmb{Our comparison uses the BUGSWARM dataset rather than the subject programs
of the original Reviser evaluation. Reproducing that evaluation is not possible
at the granularity our comparison requires: the subject applications are named,
but the specific version pairs are not, and the reported measurements cover
solver update time without stating the cost of the differencing step in isolation. Since both change set size and differencing time depend on where
the edit falls within a method rather than on the codebase alone, running our
mapper on those projects with different commits would not constitute a
comparison on the original benchmarks, and there is no published differencing
time against which to compare. We therefore re-implemented the Reviser's
differencing component and measured both approaches on the same version pairs
under identical conditions.  }

\paragraph{Construct validity.}
\cmb{We use change set size reduction as a proxy for incremental reuse
potential; actual end-to-end speedup depends on the client analysis and is not
measured here. The proxy is not hypothetical: in companion work on a pointer
analysis over several of the same subject commits, we observed that the
unmapped temporaries inflates the changed facts and shows up in parts of the
code the edit never affected, wasting code reviewer's inspection effort.
Precise mapping therefore governs both the amount of recomputation incremental
analysis performs and the fidelity of the facts it reports as changed.  }
 Furthermore, our
claim of recall\,$= 1.0$ is established by manual inspection rather
than a formal oracle, and is subject to the limitations of the
verification protocol described in
Section~\ref{sec:eval:verification}.

\section{Conclusion}
\label{sec:conclusion}

We presented a systematic, sound algorithm for mapping temporary
variables in Jimple IR across consecutive program versions---a
fundamental prerequisite for efficient incremental dataflow analysis
in the Java ecosystem. The core challenge is that Jimple
auto-generates temporary variables whose names are unstable by design:
any statement insertion or deletion causes a cascade of reindexing that
makes string-based IR matching unsuitable for this setting, as it
classifies renamed but semantically unchanged statements as modified.

Our algorithm resolves this by anchoring IR mapping to
source-line tags embedded by Soot: when source lines are confirmed as
unchanged, their associated IR statements are treated as counterparts, and
temporary variable mappings are established based on left-hand-side
alignment. We formalised three axioms characterising the properties of
java-diff-utils, Soot's line-tag generation, and Soot's deterministic IR
lowering, proved three supporting lemmas, and established a soundness
theorem guaranteeing that every produced mapping is semantically
correct. A corollary confirms that analysis facts can be safely
transferred along the map without recomputation.

Evaluated on 8 real-world Java projects with genuine edits from the BUGSWARM
dataset, our approach reduces the effective change set by an average of 82\%
compared to Reviser's graph differencing algorithm, while completing the full pipeline in
under 250\,ms.  Because the affected region an incremental analyzer must reset
and recompute is derived from the change set, a more precise change set yields
a correspondingly smaller---and likely disproportionately smaller---affected
region, compounding the practical benefit of precise IR mapping. These results
also counter the assumption that a precise change set requires more computation
than a conservative one: our approach achieves both higher precision and lower
computation time simultaneously. They demonstrate that precise, sound, and
lightweight IR mapping is not merely an implementation detail but a first-class
enabler of efficient incremental analysis. 
\cmb{As future work, we plan to measure end-to-end incremental analysis
speedup. Since the original Reviser implementation is not currently available
in working form, this requires implementing an incremental dataflow analysis
into which the mapping can be integrated, which we are pursuing separately.
}

\bibliographystyle{ACM-Reference-Format}
\bibliography{ref}



\balance
\end{document}